\documentclass[a4paper,12pt]{article}
\usepackage[spanish,USenglish]{babel}
\usepackage{times}
\usepackage{epstopdf}
\usepackage[utf8]{inputenc}
\usepackage{fancyhdr}
\usepackage{marvosym}
\usepackage{hyperref}
\usepackage{graphicx}
\usepackage{amsmath,amssymb,latexsym,amsthm,amscd}
\usepackage{calligra}
\usepackage{color}
\usepackage[T1]{fontenc}
\usepackage{lmodern}
\usepackage{fancyhdr}
\usepackage{anysize}
\marginsize{3.0 cm}{3.0cm}{1.0 cm}{2.5 cm}
\usepackage[hmarginratio=1:1,top=25mm,bottom=20mm,left=25mm,right=25mm,columnsep=20pt]{geometry}

\usepackage{amsmath}
\usepackage{amsfonts}
\usepackage{amssymb}
\usepackage{graphicx}
\newtheorem{teorema}{Teorema}
\newtheorem{definition}{Definition}
\newtheorem{lema}{Lema}
\def\R{\mathop{I\!\!R}\nolimits}
\newcommand{\Z}{{\mathbb{Z}}}

\newtheorem{com}{Remark}

\title{
	{\large\bf Shannon entropy to quantify complexity in the financial market}
	} 

\author{Alexis Rodriguez Carranza.\thanks{Mathematics institute, National University of Trujillo, Trujillo - Per\'u. ({\tt arodriguezca@unitru.edu.pe})}\and
Jos\'e Luis Ponte Bejarano.\thanks{Department of Science, Cesar Vallejo University, Trujillo - Per\'u. ({\tt jpontebe@ucvvirtual.edu.pe}).}
Juan Carlos Ponte Bejarano.\thanks{Department of Science, Cesar Vallejo University, Trujillo - Per\'u. ({\tt jcpontep@ucvvirtual.edu.pe}).}\\
Segundo Eloy Soto Abanto.\thanks{Department of Science, Cesar Vallejo University, Trujillo - Per\'u. ({\tt ssotoa@ucvvirtual.edu.pe}).}	
}
\begin{document}
 \maketitle
 
 




%

%
\selectlanguage{English}
\begin{center}
	{\bf Abstract}
\end{center}
{\it \noindent In this paper we study the complexity in the information traffic that occurs in the peruvian financial market, using the Shannon entropy. Different series of prices of shares traded on the Lima stock exchange are used to reconstruct the unknown dynamics. We present numerical simulations on the reconstructed dynamics and we calculate the Shannon entropy to measure its complexity.}

 
 
\selectlanguage{spanish}
%


\section{Introduction}
In a organized and developed society, the economic aspect plays a fundamental role for the well-being of the members of society, which is why understanding and knowing the dynamics of the financial market allows sustainable social growth over time. 
The crisis that has shaken the world economy should raise questions for economists about the approach used to analyze economic phenomena. The classical economic models that are used in the asset market, see [8], simulate the volatile behavior of exchange rates such as the prices of financial evaluations negotiated in efficient markets, but the current exchange rate contains the information available instantaneously and the observed changes reflect the effect of new events that are unpredictable by definition. 

Information theory, particularly Shannon entropy, has been used in this field to understand financial market behavior. For example, Chen J. discusses the similarity between information theory concepts and the economic value of information with respect to markets [1], and uses Shannon's entropy to explain most of the empirical evidence about behaviors of the market, determining the value of the information according to the number of people who know it [2].
Maasoumi and Racine applied entropy to find non-linear dependencies of stock price returns and their predictions [7]. Garca, Cruz and Venegas use
Shannon entropy to propose a market efficiency measure to be applied to different capital markets: DJIA, S \& P500, FTSE100 and IPC [5].

These investigations show that Shannon's entropy has been applied to different fields of finance in an optimal way. In our case we will use it to determine the number of bits necessary to obtain information about what happens with the evolution of the behavior of prices in the financial market. We will do our research on a couple of price time series: stocks of the companies TelfBC and Credicorp to determine the complexity in the financial market. For this, it will be necessary to determine clues about the presence of a fractal attractor set in the dynamics of the series. Figure 1 shows the evolution of the prices of these shares. 
\begin{figure}[h]
  \begin{center}
 \scalebox{0.35}{\includegraphics{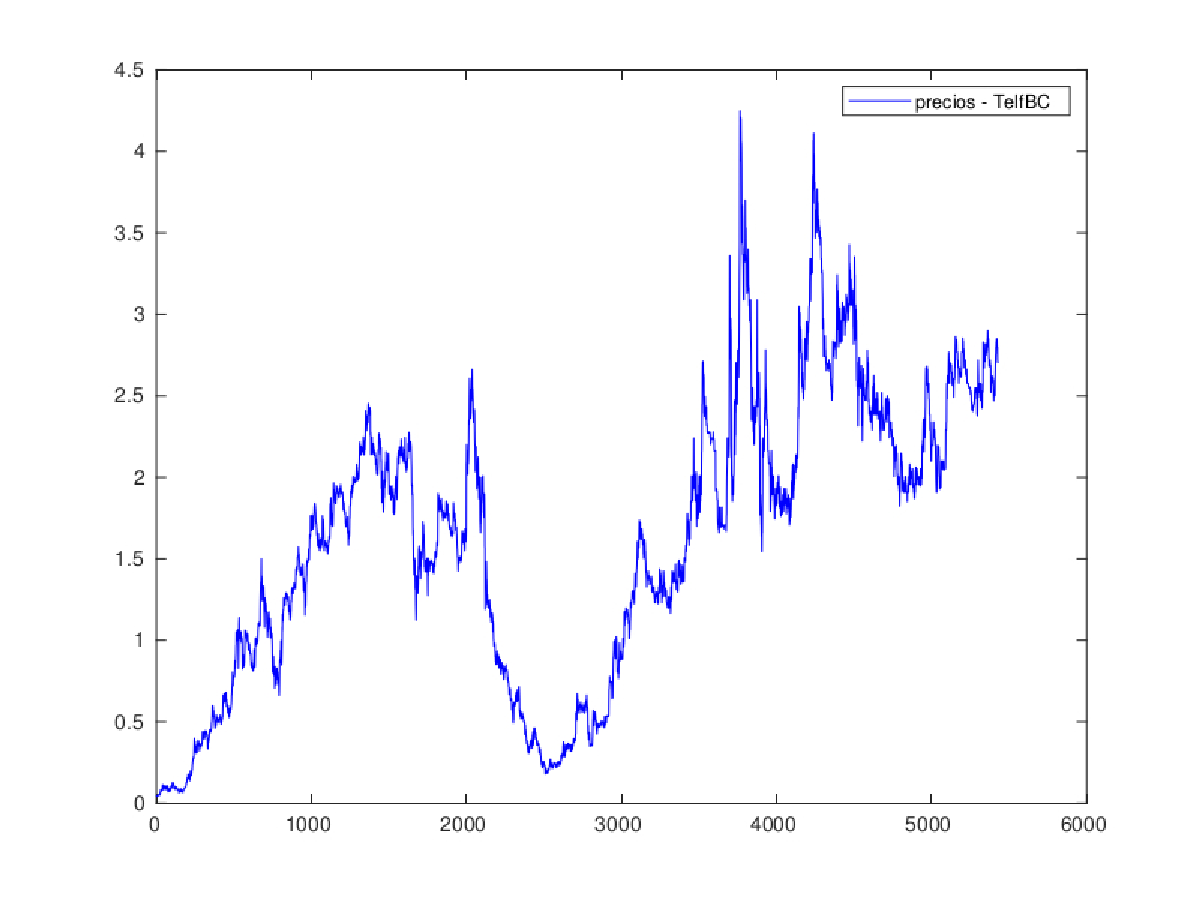}}
 \scalebox{0.35}{\includegraphics{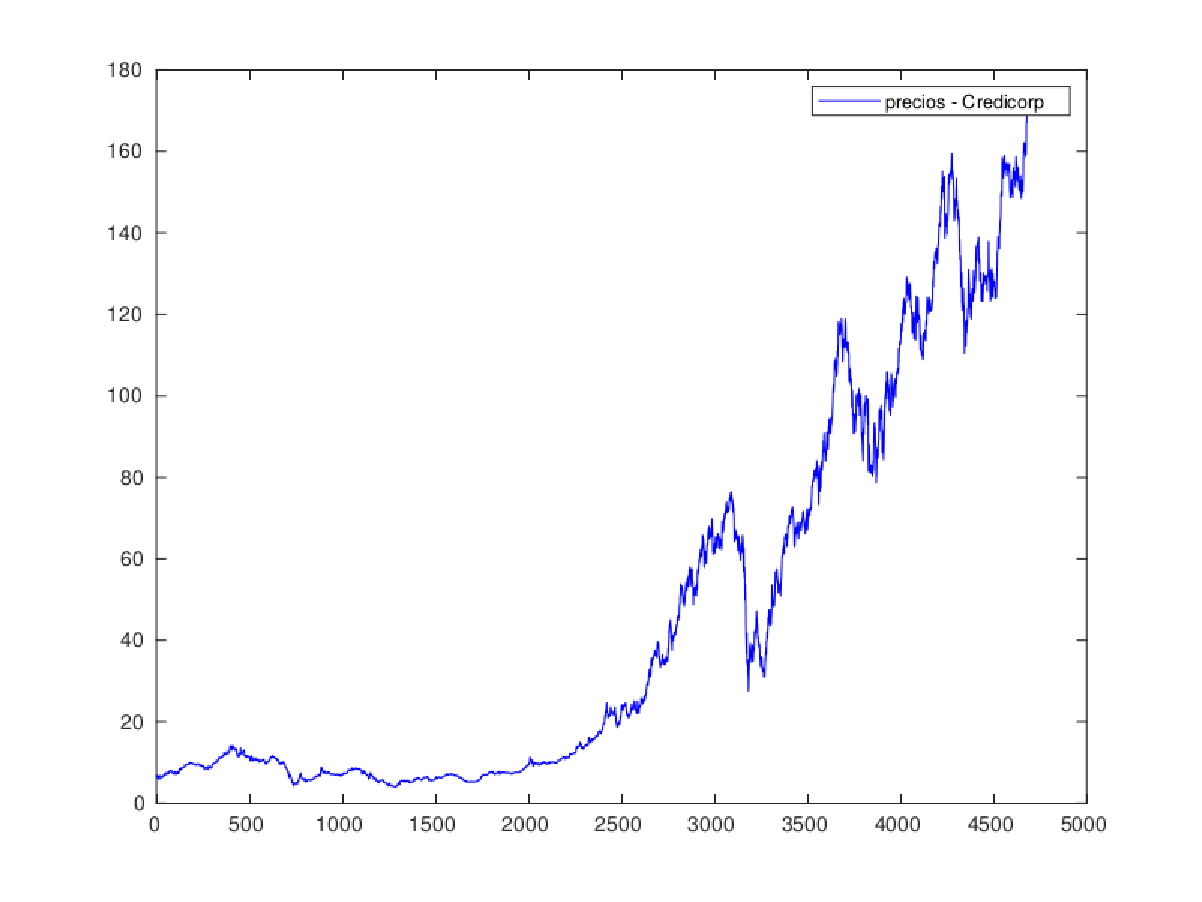}}
  \end{center}
 \caption{Series of prices of the shares of the companies TelfBC and Credicorp.}
 \label{precios}
 \end{figure}

\section{Reconstruction theorems}
\begin{figure}[h]
  \begin{center}
 \scalebox{0.35}{\includegraphics{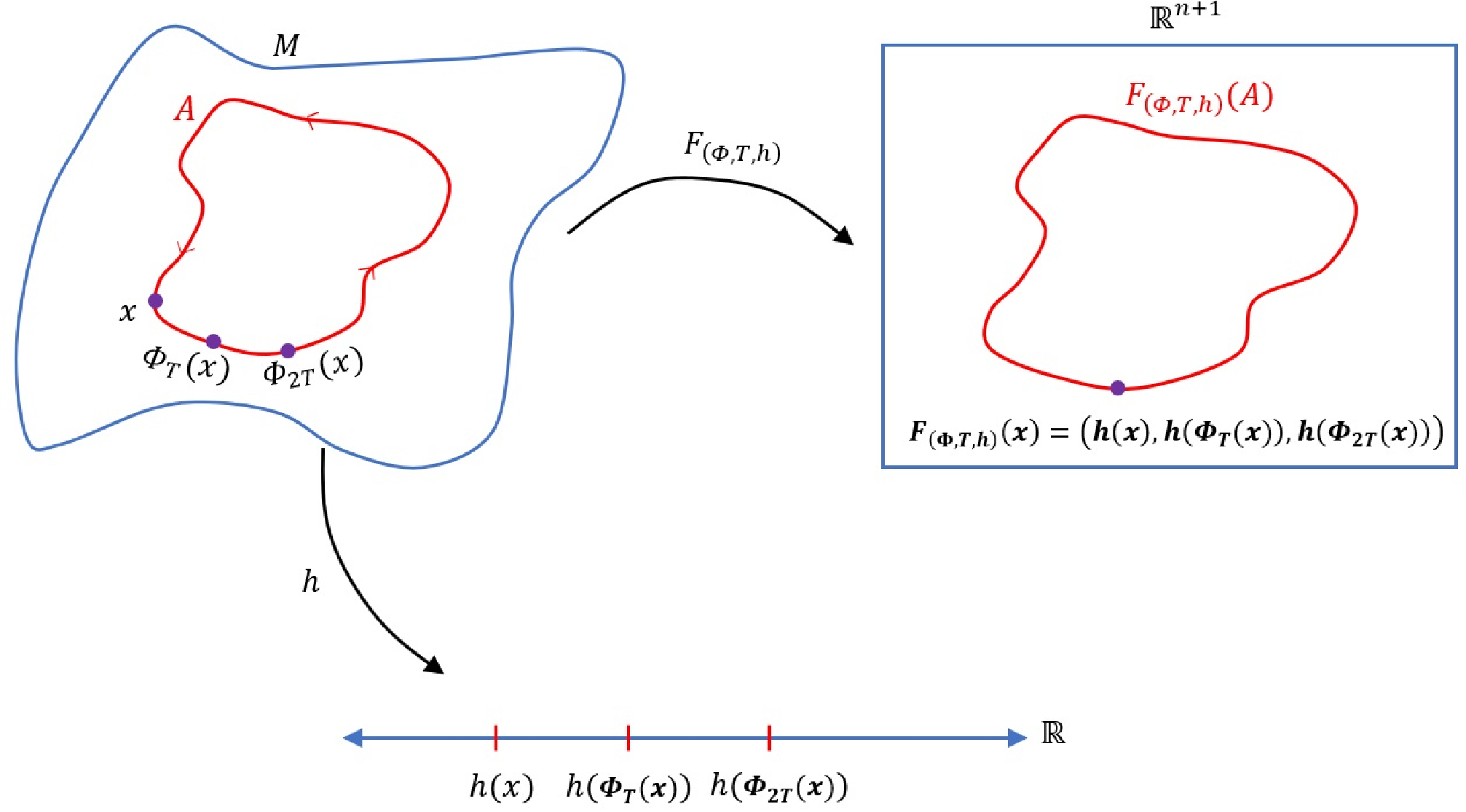}}
 \end{center}
 \caption{Application of delay coordinates $F_{(\Phi,T,h)}$; and reconstruction of the attractor $A$ using the application $F_{(\Phi,T,h)}$ in $\R^{n+1}$ with $n=2$.}
 \label{construccion3}
 \end{figure}

\begin{definition}[time serie]\label{serietemporal}
Given a dynamic system with flow $\{\phi_{t}\}_{t\in\R}$ in a phase space $H$. A time series $s:\R \longrightarrow\R$ is defined as the values that a function takes, called the observation function, $F:H\longrightarrow\R$ given by $s(t)=F(\phi_{t}(x_{0}))$, $t\in\R,  x_{0}\in H$.
\end{definition}
For the reconstruction of the hidden attractor in the time series, Takens [11] uses the information of the dynamic system contained in the time series. For such a reconstruction, the delay coordinates are defined, with which a single temporal observation is needed. 
\begin{definition}
Be $\Phi$ the flow of a dynamic system over a differentiable manifold $M$ of at least class $C^{1}$, $T$ a positive integer (called delay) and  $h:M\rightarrow\R $ a class function $C^{1}$. The delay coordinate application is defined  $F_{(\Phi,T,h)}:M\rightarrow\R^{n+1}$ by:
$$F_{(\Phi,T,h)}(x)=\left(h(x),h(\Phi_{T}(x)),h(\Phi_{2T}(x)),\ldots,h(\Phi_{nT}(x))  \right) $$
\end{definition}
A geometric idea of the application of delay coordinates $F_{(\Phi,T,h)}$  is shown in figure \ref{construccion3}.

\begin{teorema}[Takens embedding Theorem \cite{Takens}]\label{takens}
Be $M$ a differentiable variety, at least of class $C^{2}$, compact of dimension $m$; $\{\Phi_{k}\}_{k\in\Z}$ the flow of a discrete dynamic system over $M$ with $\Phi_{k}:M\rightarrow M$ a diffeomorphism of class  $C^{2}$; and $h:M\rightarrow\R$ a function of class $C^{2}$. So it is a generic property that the application $F_{(\Phi,h)}:M\rightarrow\R^{2m+1}$ defined by 
$$F_{(\Phi,h)}(x)=\left(h(x),h(\Phi_{k}(x)),h(\Phi_{2k}(x)),\ldots,h(\Phi_{2mk}(x))  \right)$$
is a embedding of $M$.
\end{teorema}

Taken's Theorem tells us that if we have a dynamic system that depends on $m$ variables, and if we choose an observation function $h$, for this system with those measurements we construct the vectors: 
$$\left(h(x),h(\Phi_{k}(x)),h(\Phi_{2k}(x)),\ldots,h(\Phi_{2mk}(x))  \right)$$
$$\left(h(\Phi_{k}(x)),h(\Phi_{2k}(x)),\ldots,h(\Phi_{2mk}(x)),h(\Phi_{2mk+1}(x))  \right)$$
$$\vdots$$
we can have a copy of the original dynamic system.\\
This result is important, since we can observe one of the variables of the system over time and by choosing the dimension of the reconstruction vectors properly, we can understand the evolution of the system of m variables, as illustrated in Figure 3. 
\begin{figure}[h]
  \begin{center}
 \scalebox{0.45}{\includegraphics{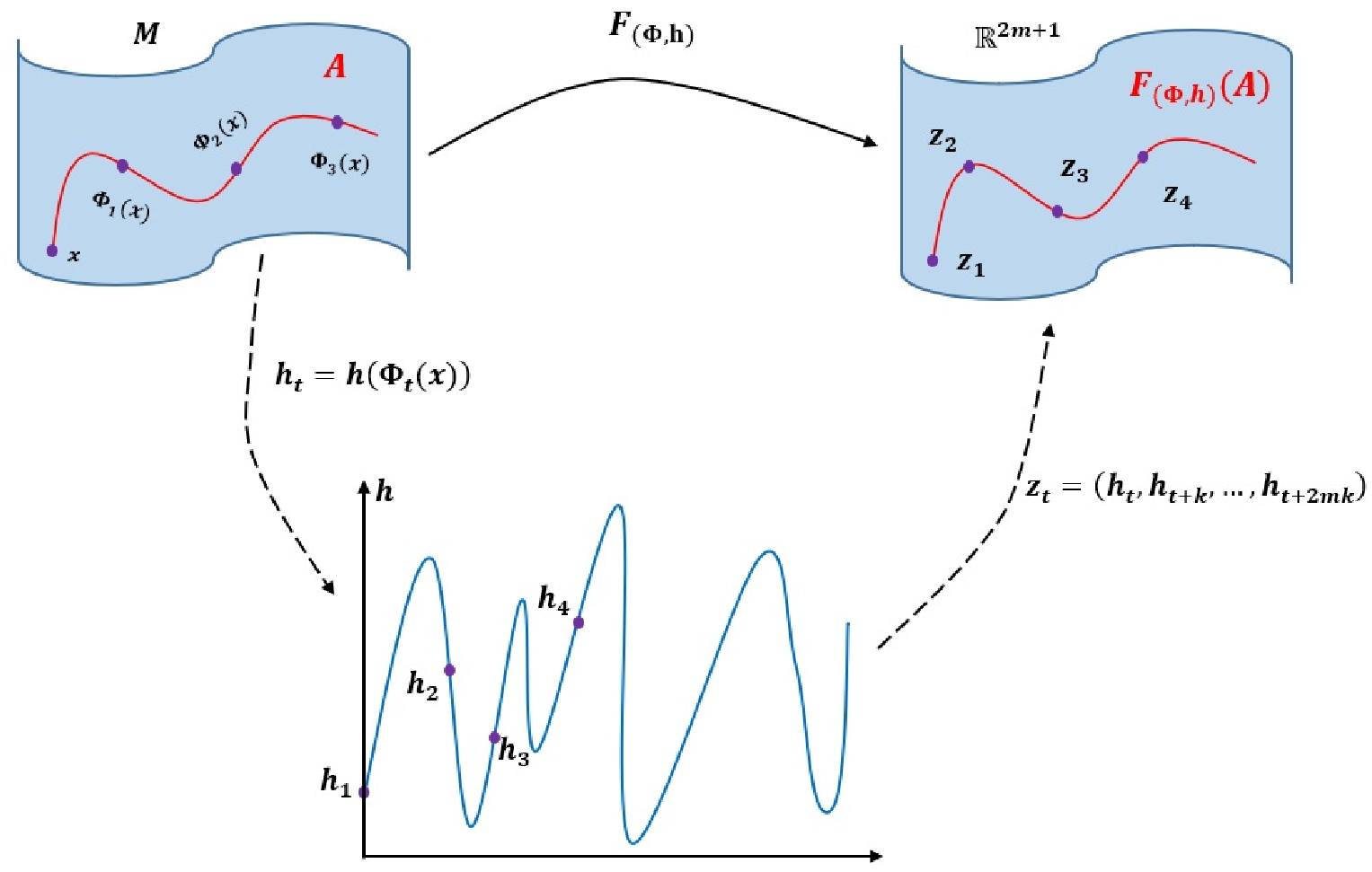}}
  \end{center}
 \caption{Takens embedding Theorem. Reconstruction of the phase space $ M $ and the attractor $ A $ via the application of delay coordinates  $F_{(\Phi,h)}$.}
 \label{FiguraTakens}
 \end{figure}

The final generalization, used in this article, was given by Tim Sauer, James A. Yorke, and Martin Casdagli [9]. They propose that it is possible to have an injective copy of the attractor via the application of delay coordinates with A being a fractal set. Some results and definitions are necessary. 

\begin{lema}\cite{Sauer}\label{lema12}
Let $A$ a compact subset of $\R^{k}$ and $F_{0},F_{1},\ldots,F_{t}$ Lipschitz applications of $A$ on $\R^{n}$. For each integer $r\geq 0$, be $S_{r}$ the set of pairs $x\neq y$ on $A$ for which the matrix $n\times t$
$$M_{xy}=\left[ F_{1}(x)-F_{1}(y),\ldots,F_{t}(x)-F_{t}(y) \right]$$
has rank $r$, and $d_{r}=\dim_{B}(\overline{S_{r}})$. For each $\alpha\in\R^{t}$, with $\alpha =(\alpha^{1},\dots,\alpha^{t})$, define $F_{\alpha}=F_{0}+\displaystyle\sum_{i=1}^{t}\alpha^{i} F_{i}:A\rightarrow\R^{n}$. Then, for $\alpha\in\R^{t}$ out of a subset of zero measure of $\R^{t}$, it is true that yes $d_{r}<r$ for all integers  $r\geq 0$, then the application $F_{\alpha}$ it is injective.
\end{lema}

\begin{definition}
A Borel subset A of a normed vector space V is prevalent if there exists a subspace, E of V, of finite dimension such that for each $v \in V, v + e \in A$ for almost everything (Lebesgue measure) $e \in E.$
\end{definition}
We will use the term for almost every application if the set of such functions is prevalent. 
\begin{definition}
Let $I_{n}$ be the identity matrix of $n\times n$ and $(\cdot,\cdot)$ denotes the greatest common divisor. Let's use the convention that $(p,0)=0$. For the integers $p>q\geq 0$ let's define the matrix $p\times (p-(p,q))$ by: 
\begin{eqnarray*}
C_{pq}&=&\left[
\begin{array}{lll}
 & I_{p-(p,q)} &\\
 \\
-I_{(p,q)}&\quad\cdots &-I_{(p,q)}\\
\end{array}
\right]
\end{eqnarray*}\label{DefE1}
Let's define $C^{\infty}_{pq}$ a matrix $\infty\times (p-(p,q))$ formed by repeating the block $C_{pq}$ vertically. And for a positive integer $w$ let's define $C^{w}_{pq}$ the matrix formed by the $w$ top rows of $C^{\infty}_{pq}$.
\end{definition}
\begin{com}
If $p\leq n$, of the definition \ref{DefE1} it is observed that rank $C^{n}_{pq}=p-(p,q)$; and then rank $C^{n}_{pq}=\min\{n,p-(p,q) \}$. From this it follows that: 
\begin{itemize}
\item[a)] rank $C^{n}_{p0}\geq \min\{n,p\}=p$ and
\item[b)] rank $C^{n}_{pq}\geq \min\{n,p/2 \}=p/2$.
\end{itemize}
\end{com}

\begin{teorema}[Fractal Delay Embedding Prevalence Theorem \cite{Sauer}]\label{sauerr}
Be $\Phi$ the flow of a dynamic system over an open subset $U$ of $\R^{k}$, and $A$ a compact subset of $U$ of $\dim_{B}(A)=d$. Be $n>2d$ an integer and $T>0$. Let's assume that $A$ contains only a finite number of equilibrium points; does not contain periodic orbits of $\Phi$ of period $T$ or $2T$ y contains a finite number of periodic orbits of $\Phi$ of period $3T, 4T, \ldots, nT$. So for almost all (in the prevalente sense) function  $h$ of class $C^{1}$ on $U$, the delay coordinate application $F_{(\Phi,T,h)}:U\rightarrow\R^{n}$, defined by:
$$F_{(\Phi,T,h)}(x)=\left(h(x),h(\Phi_{T}(x)),h(\Phi_{2T}(x)),\ldots,h(\Phi_{(n-1)T}(x))  \right), $$
it is injective on $A$.
\end{teorema}
\begin{proof}
Be $\{h_{i}\}_{i=1}^{t}$ a basis for the polynomials of $k$ variables of lesser degree and equal to $2n$. For $i=1,\dots, t$ let's define
\begin{eqnarray*}
F_{i}(x)&=&\left[
\begin{array}{l}
 h_{i}(x)\\
 h_{i}(\Phi_{T}(x))\\
 h_{i}(\Phi_{2T}(x))\\
\quad \vdots\\
h_{i}(\Phi_{(n-1)T}(x))\\
\end{array}
\right]
\end{eqnarray*}
For each $\alpha\in\R^{t}$, with $\alpha=(\alpha^{1},\dots,\alpha^{t})$, let's define:
$$h_{\alpha}=\sum_{i=1}^{t}\alpha^{i}h_{i}:\R^{k}\rightarrow\R.$$
Then,
$$F_{(\Phi,T,h_{\alpha})}=\sum_{i=1}^{t}F_{i}:\R^{k}\rightarrow\R^{n}.$$
To use the lemma \ref{lema12} we need to determine, for each $(x,y)\in A\times A$ with $x\neq y$, the rank of the matrix: 
$$M_{xy}=[F_{1}(x)-F_{1}(y),\dots,F_{t}(x)-F_{t}(y)]_{n\times t},$$
which can be expressed as:
\begin{eqnarray*}
M_{xy}&=&\left[
\begin{array}{lll}
 h_{1}(x)-h_{1}(y)&\cdots &h_{t}(x)-h_{t}(y)\\
 h_{1}(\Phi_{T}(x))-h_{1}(\Phi_{T}(y))&\cdots &h_{t}(\Phi_{T}(x))-h_{t}(\Phi_{T}(y)) \\
\qquad\qquad \vdots &\ddots & \qquad\qquad \vdots\\
h_{1}(\Phi_{(n-1)T}(x))-h_{1}(\Phi_{(n-1)T}(y))&\cdots &h_{t}(\Phi_{(n-1)T}(x))-h_{t}(\Phi_{(n-1)T}(y)) \\
\end{array}
\right]=J\cdot H,
\end{eqnarray*}
where, 
\begin{eqnarray*}
H &=&\left[
\begin{array}{llll}
 h_{1}(z_{1})&\cdots &h_{t}(z_{1})\\
 h_{1}(\Phi_{T}(z_{2}))&\cdots & h_{t}(\Phi_{T}(z_{2})) \\
\qquad \vdots &\ddots & \qquad \vdots\\
h_{1}(\Phi_{(n-1)T}(z_{q}))&\cdots &h_{t}(\Phi_{(n-1)T}(z_{q})) \\
\end{array}
\right]_{q\times t},
\end{eqnarray*}
$q\leq 2n$; the $z_{j}$, $j=1,\dots q$, are different and $J$ is a matrix of $n\times q$ whose rows consist of $1, 0$ y $-1$.\\
For each integer $p\leq n$ positive let us denote by $A_{p}$ the set of periodic period points $p$ of $\Phi_{T}$ found in $A$, that is to say, $$A_{p}=\{x\in A:\Phi_{T}^{p}(x)=x\}.$$
From the statement of the theorem, as $A$ contains a finite number of periodic period points $p\leq n$  of the flow $\Phi_{T}$, then $$\dim_{B}(A_{p})<p/2.$$
Now we will divide the study of the rank of $M_{xy}$ in three cases:\\
Case 1: The points $x$ and $y$ they are not periodicals $\leq n$.\\
In this case $J$ is an upper or lower triangular matrix and rank $J=n$. Later, $$\mbox{rank }M_{xy}=\mbox{rank }(J\cdot H)=n.$$
Let's define the set $$S_{n}=\{(x,y)\in A\times A: x\neq y\mbox{ con rank }M_{xy}=n \},$$
from where you have:
$$\dim_{B}(S_{n})\leq 2d<n.$$
So, by the lemma \ref{lema12} we conclude that for almost everything $\alpha\in\R^{t}$, the application $F_{(\Phi,T,h_{\alpha})}$ it is injective on $A$.\\
Case 2: The points $x$ and $y$ are in different periodic orbits of period $\leq n$.\\
Let's assume that $p$ and $q$ are minimal positive integers such that $\Phi_{T}^{p}(x)=x$, $\Phi_{T}^{q}(y)=y$ and $1\leq q\leq p\leq n$. In this case, the matrix $J$ contains a copy of $C_{p0}^{n}$. Then, $$\mbox{rank }M_{xy}=\mbox{rank }(J\cdot H)=\mbox{rank }J\geq \mbox{rank }C^{n}_{p0}>2\dim_{B}(A_{p}).$$
Let's define the set $S_{r}$, with $x$ and $y$ the points dealt in the case 2, by: $$S_{r}=\{ (x,y)\in A\times A: x\neq y \mbox{ con rank }M_{xy}=r\},$$
from where:
$$\dim_{B}(S_{r})\leq 2\dim_{B}(A_{p})<\mbox{rank }M_{xy}=r.$$
Then, for the lemma \ref{lema12}, we conclude that for almost everything $\alpha\in\R^{t}$ the application $F_{(\Phi,T,h_{\alpha})}$ it is injective on $A$.\\
Case 3: The points $x$ and $y$ are in the same periodic orbit of period $\leq n$.\\
Let's assume that $p$ and $q$ are minimal positive integers such that $\Phi_{T}^{p}(x)=x$, $\Phi_{T}^{q}(x)=y$ and $1\leq q< p\leq n$. Given that $x$ and $y$ lie in the same periodic orbit, the column space (the set of linearly independent columns) of  $J$ contains the column space of $C^{n}_{pq}$. Thus, $$\mbox{rank }M_{xy}=\mbox{rank }(J\cdot H)=\mbox{ rank }J>\mbox{rank }C^{n}_{pq}>\dim_{B}(A_{p}).$$
Let's define the set $S_{r}$, with $x$ and $y$ the points dealt in the case 3, by: $$S_{r}=\{ (x,y)\in A\times A: x\neq y \mbox{ con rank }M_{xy}=r\}.$$
As $x$ and $y$ are in the same periodic orbit, $$\dim_{B}(S_{r})=\dim_{B}(A_{p})<\mbox{rank }M_{xy}=r.$$
Then, for the lemma \ref{lema12}, we conclude that for almost everything $\alpha\in\R^{t}$ the application $F_{(\Phi,T,h_{\alpha})}$ it is injective on $A$. This concludes the demonstration.
\end{proof}
We should mention that the Theorem \ref{sauerr}  does not give an estimate on the smallest dimension for which almost every application of delay coordinates is injective. However, there are numerical algorithms that allow estimating the mergulho dimension and the delay time in the reconstructions. These are the mutual information and the false neighbor method, which we mention in the next section.

\section{Mutual information and false neighbors}
\subsection{Mutual information}
\begin{definition}
Let $X$ and $Y$ two discrete random variables with probability distributions $p(x)$ and $p(y)$, respectively. Mutual information between variables $X$ and $Y$ is defined by:
$$I(X,Y)=\sum_{x}\sum_{y}p(x,y)\log_{2}\frac{p(x,y)}{p(x)p(y)},$$
where $p(x,y)$ represents the joint probability.
\end{definition}
If we consider that the random variable $X$ take the values $x_{1},x_{2},\ldots ,x_{N}$ So what $Y$ take the values $x_{1+T}, x_{2+T},\ldots ,x_{N+T}$, the mutual information for the time series is expressed as a function of $T$:
$$I(T)=\sum_{t=1}^{N}P(x_{t},x_{t+T})\log_{2}\frac{P(x_{t},x_{t+T})}{P(x_{t})P(x_{t+T})},$$
where $P(x_{t})$ is the probability that $X$ take a value $x_{t}$, $P(x_{t+T})$ is the probability that $Y$ take a value $x_{t+T}$. And $P(x_{t},x_{t+T})$ is the probability that $X$ take a value $x_{t}$ and $Y$ take a value $x_{t+T}$.\\
\begin{com}
We comment the following:
\begin{itemize}
\item[a)] The mutual information acts as the nonlinear autocorrelation function that indicates how, in a nonlinear way, the measurements at different times are connected on average over all measurements.
\item[b)]Mutual information between $x_{t}$ y $x_{t+T}$ quantifies the information you have about the state $x_{t+T}$ assuming we have knowledge of the state $x_{t}$.
\end{itemize} 
\end{com}
Fraser and Swinney \cite{Fraser} propose to use the first minimum of the mutual information between $x_{t}$ and $x_{t+T}$ as the optimal delay time. The idea is that a suitable delay time has to be large enough so that the information available in time $t+T$ is significantly different from the information over time $t$, but, not too much so that this information is not lost.\\
For the calculation of the mutual information of a time series, $\{x_{1},x_{2},\ldots, x_{N} \}$, the first step is to find the maximum, $x_{\max}=\max\{x_{1},x_{2},\ldots, x_{N} \}$ and the minimum, $x_{\min}=\min\{x_{1},x_{2},\ldots, x_{N} \}$ of the series values. Then, the value $\vert x_{\max}-x_{\min}\vert$ is divided into $j$ intervals of the same size. Finally the expression is calculated
$$I(T)=\sum_{h=1}^{j}\sum_{k=1}^{j}P_{h,k}(T)\ln \frac{P_{h,k}(T)}{P_{h}P_{k}}$$
where $P_{h}$ and $P_{k}$ denote the probabilities that the variables take the values in the $h-$esimo and $k-$esimo interval, respectively, and $P_{h,k}(T)$ is the joint probability that $x_{t}$ is in the interval $h$ and $x_{t+T}$ is in the interval $k$.\\
The first local minimum of $I(T)$ indicates the largest amount of information we can have from the state $x_{t}$ in order to determine the status $x_{t+T}$.\\
To obtain the dimension of the space where the attractor is rebuilt, called the embedding dimension, let's see the false neighbors method.
\subsection{False neighbors}
The technique of \textit{False neighbors} was introduced by Kennel \cite{Kennel}. It is an efficient tool to determine the smallest required reconstruction dimension, that is, the embedding dimension. By reconstruction theorems, we are only guaranteed that for an adequate reconstruction dimension, say $m$, a copy of the attractor can be obtained at $\R^{m}$. In addition, for $n>m$ the same thing happens. In this sense, the false neighbors method can serve as an optimization procedure trying to have the lowest value of $m$, in such a way as to have the reconstruction of the adequate phase space.\\
The idea of false neighbors is based on the following geometric intuition:\\
We observe that the dimension of an attractor expresses the degrees of freedom that are needed to extend. Likewise, if the dimension of the reconstruction space were less than the dimension of the attractor, it would not have sufficient degrees of freedom for it, being reduced. Then, nearby points in that state can be mistaken as true neighbors. By increasing the reconstruction dimension, the attractor will deform and false neighbors will no longer be neighbors. The following algorithm estimates the value of $m$ for which a fraction of false neighbors is null. 

To calculate the fraction of false neighbors, the following algorithm is used: Given a point $X_{t}^{(m)}=\left(x_{t},x_{t+\tau},x_{t+2\tau},\ldots,x_{t+(m-1)\tau} \right)$ in the reconstruction space $m-$dimensional, we have to find a neighbor $X_{i}^{(m)}$ such that $\Vert X_{i}^{(m)}-X_{t}^{(m)} \Vert<\varepsilon$, donde $\varepsilon$ is a small constant. So, we calculate the normalized distance $R_{i}$ between the coordinates in $\R^{m+1}$ of the points $X_{i}^{(m+1)}=(X_{i}^{(m)},x_{i+m\tau})$ y $X_{t}^{(m+1)}=(X_{t}^{(m)},x_{t+m\tau})$ according to the following equation:
\begin{equation}
 R_{i}=\displaystyle\frac{\vert x_{i+m\tau}-x_{t+mr} \vert}{\Vert X_{i}^{(m)}-X_{t}^{(m)} \Vert}
\end{equation}
If $R_{i}$ is greater than a given referential value $R_{tol}$, then $X_{i}^{(m)}$ is marked as a false neighbor of $X_{t}^{(m)}$. Equation (1) has to be calculated for the whole series and for several values of $m=1,2,\ldots ,$ until the fraction of points for which $R_{i}>R_{tol}$ be despicable. In accordance with Kennel \cite{Kennel}, $R_{tol}=10$ It has proven to be a good choice for a considerable number of series. A formal mathematical proof of this fact is non-existent. 

Next we will give a brief summary about how you can think about the concept of the dimension of an object \cite{Theiler}. \\
We will use the following ideas to define the information dimension, which will be useful to detect the fractal structure.
\section{Shannon entropy and information dimension}
\begin{figure}[h]
  \begin{center}
 \scalebox{0.3}{\includegraphics{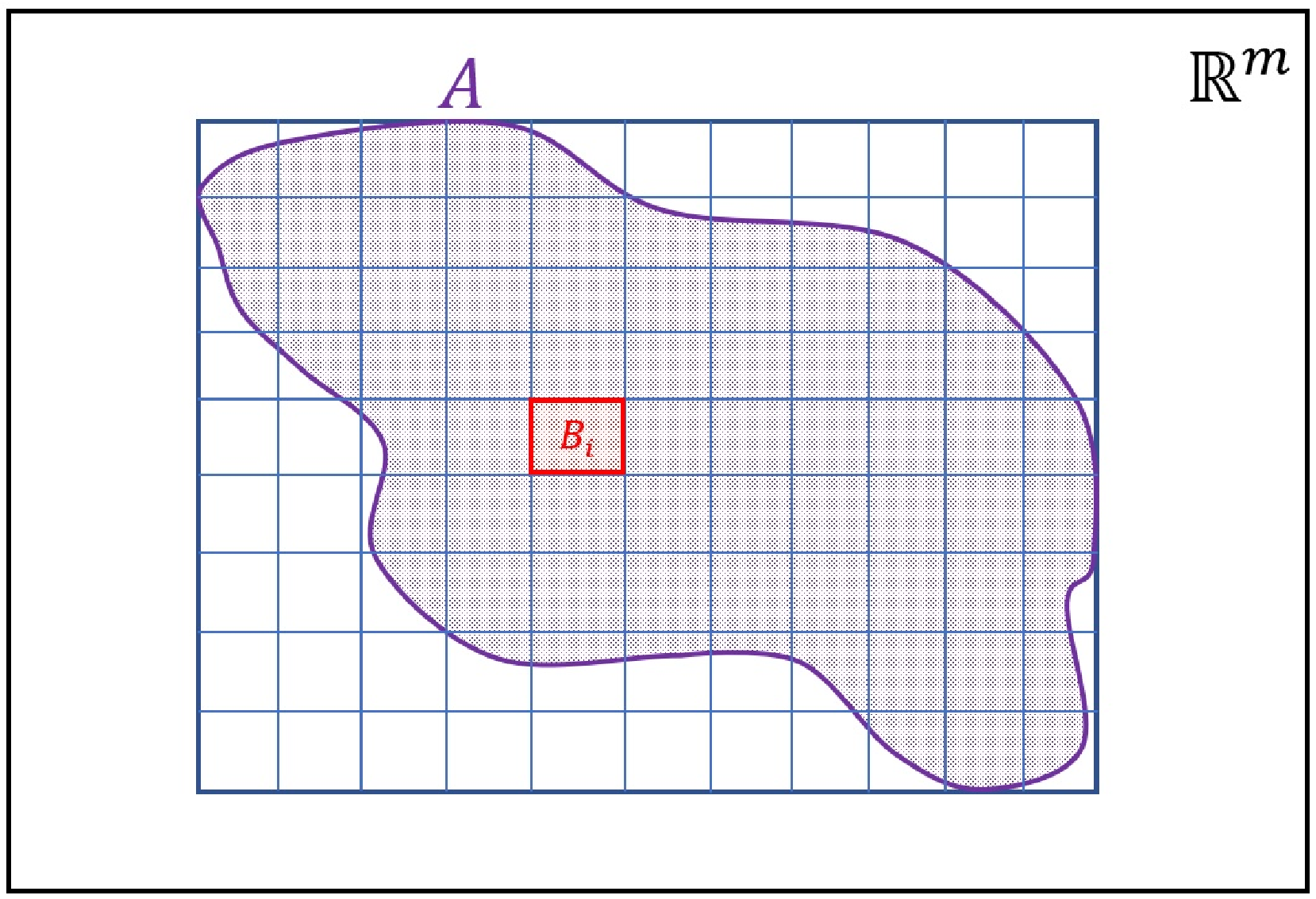}}
  \end{center}
 \caption{Partition of the set $ A $ in $ n $ boxes $ B_ {i} $ of diameter $r<1$.}
 \label{particion}
 \end{figure}

A way of thinking about the dimension of a set $A$, is in terms of how many real numbers are needed to specify the position of a point in that set. For example, the position of a point on the line is determined by one number, the position on a plane by two. Now, the idea is to understand this concept for more complicated sets than lines, planes, volumes, etc.\\
We will need the following definitions:
\begin{definition}
Be $A$ a nonempty set of $R^{m}$. The information dimension of set A is given by:
$$D_{I}=\lim_{r\rightarrow 0}\frac{-S(r)}{\log_{2}(r)}$$
where $S(r)$ is the information (in bits) needed to specify the position of a point on the set $A$ with a precision $r$.
\end{definition}
The calculation of $S(r)$ was given by Claude Shannon \cite{Shannon}.
\begin{definition}
Be $X$ a discrete random variable defined in a set A, with $n-$events whose probabilities of occurrence are $p_{1},p_{2},\ldots,p_{n}$. The Shannon entropy is defined from $X$, by:
$$H(X)=-\sum_{i=1}^{n}p_{i}\log_{2}(p_{i})$$
\end{definition}
\begin{com}
For clarity of Shannon's concept of entropy, the following observations should be kept in mind:
\begin{itemize}
\item[1)] Shannon entropy does not depend on the values that the random variable can take $X$, but only of probabilities. Therefore, the Shannon entropy is a function of a certain probability distribution. $p=(p_{1},p_{2},\dots,p_{n})$, so this is usually denoted by $H(p)$.
\item[2)] Since the base of the logarithm is $2$, Shannon entropy is measured in bits. This is an average measure of the uncertainty of the random variable, that is, it is the average number of bits required to describe the random variable $X$. This means that the Shannon entropy $H(p)$ represents the average amount of information to specify the position of a point in $A$.
\end{itemize}
\end{com}
\begin{teorema}\label{dim}
Let A be a nonempty subset of $\R^m$ and $X$ a discrete random variable defined in $A$ with $n-$events, whose probabilities of occurrence are $p_{1},p_{2},\ldots,p_{n}$. Then, the information dimension of the set $ A $ is given by: 
$$D_{I}=\lim_{r\rightarrow 0}\frac{\displaystyle\sum_{i=1}^{n}p_{i}\log_{2}(p_{i})}{\log_{2}(r)},$$
where $r$ represents the precision with which a point is specified in $A$.
\end{teorema}
\begin{proof}
For the set $ A $ let us consider a partition of it in $n$-boxes $B_{i}$ of diameter $r$. Then, the probability that a point in the set $A$ be in the box $B_{i}$ is given by $p_{i}=\displaystyle\frac{\mu(B_{i})}{\mu(A)}$. Using Shannon's entropy, the information needed to specify a point in the set $ A $ with precision $ r $ is given by:
$$S(r)=-\sum_{i=1}^{n}p_{i}\log_{2}(p_{i})$$
This relationship leads directly to an expression for the information dimension of the set $ A $:
\begin{eqnarray*}
D_{I}&=&\lim_{r\rightarrow 0}\frac{-S(r)}{\log_{2}(r)}\\
D_{I}&=&\displaystyle\lim_{r\rightarrow 0}\frac{\displaystyle\sum_{i=1}^{n}p_{i}\log_{2}(p_{i})}{\log_{2}(r)}
\end{eqnarray*}
\end{proof}
This concludes the proof.\\
In the figure \ref{particion} a partition of the set $ A $ is observed, for which
$$p_{i}=\displaystyle\frac{\mbox{ $N^{0}$ of points in } B_{i} \mbox{ of } A}{\mbox{ $N^{0}$ of points in } A}$$

\section{Applications to the financial market}
In this section we will apply the techniques elaborated in previous sections to reconstruct the hidden attractor in the price series of the companies TelfBC and Credicorp. In addition, the Shannon entropy of the reconstructed dynamics will be determined, and then the information dimension of this will be calculated. The numerical value of the information dimension will indicate the complexity in the financial market. Also, it will tell us that the reconstructed attractor is a fractal set. \\
The first series of prices studied was from the company TelfBC, with prices from 01/02/1992 to 11/26/2014, a total of 21 years of daily observations. And the second series of prices was from the Credicorp company, with prices from 10/25/1995 to 11/26/2014, a total of 19 years of daily observations.
\subsection{Reconstruction of attractors}
\begin{figure}[h]
  \begin{center}
  \scalebox{0.37}{\includegraphics{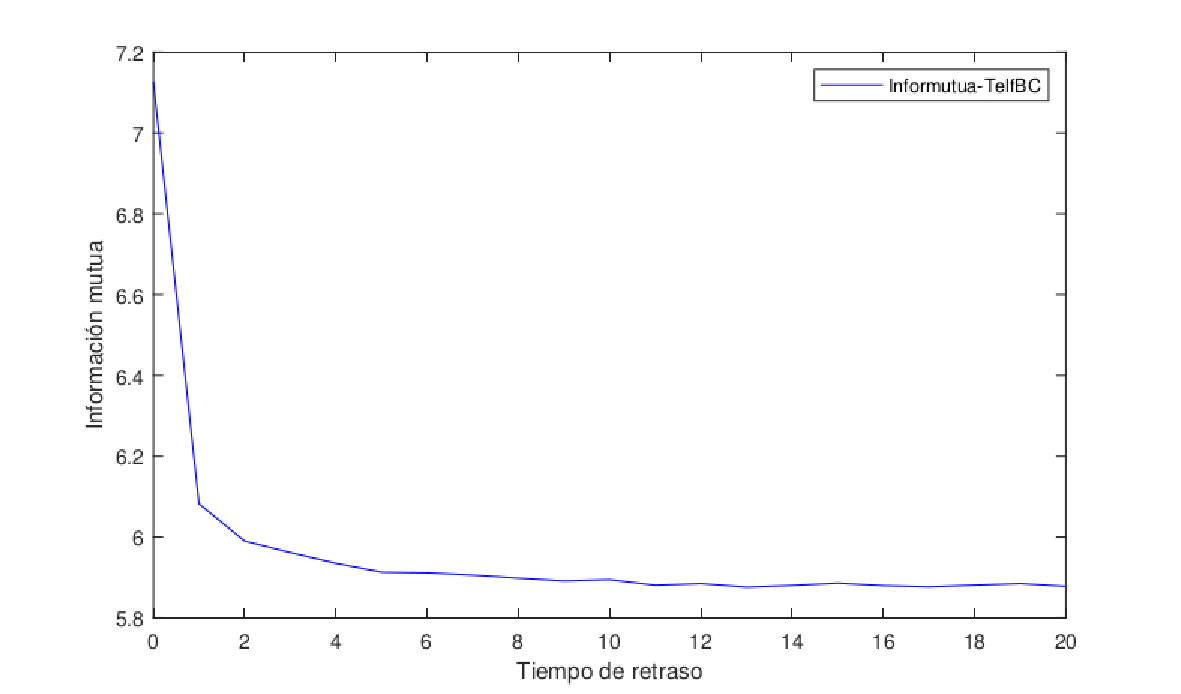}}
 \scalebox{0.37}{\includegraphics{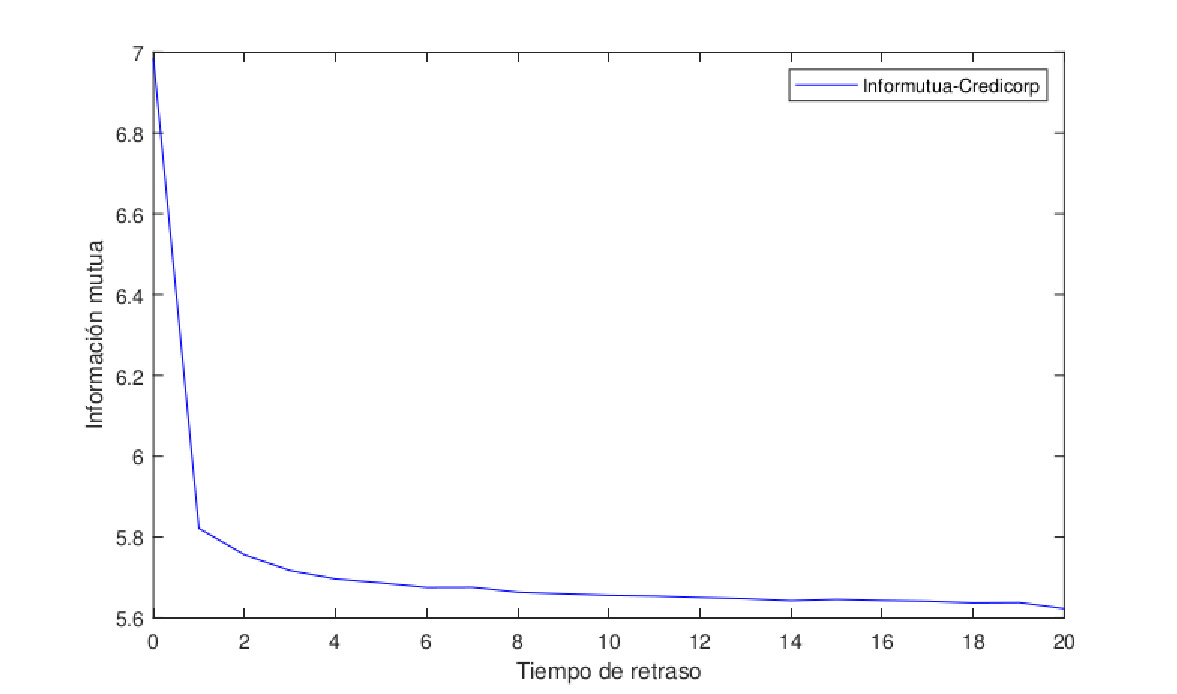}}
  \end{center}
 \caption{Delay time $ T = $ 9 and $ T = $ 6, corresponding to the price series of the companies TelfBC and Credicorp.}
 \label{delay}
 \end{figure}

\begin{figure}[h]
  \begin{center}
  \scalebox{0.37}{\includegraphics{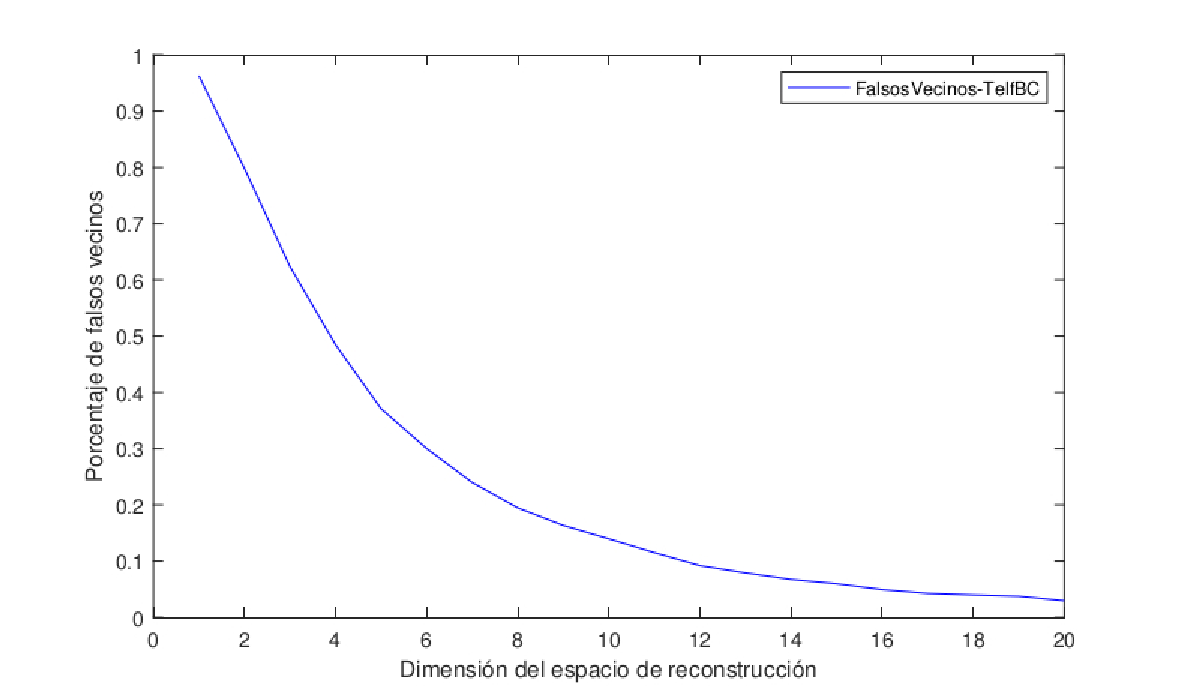}}
 \scalebox{0.37}{\includegraphics{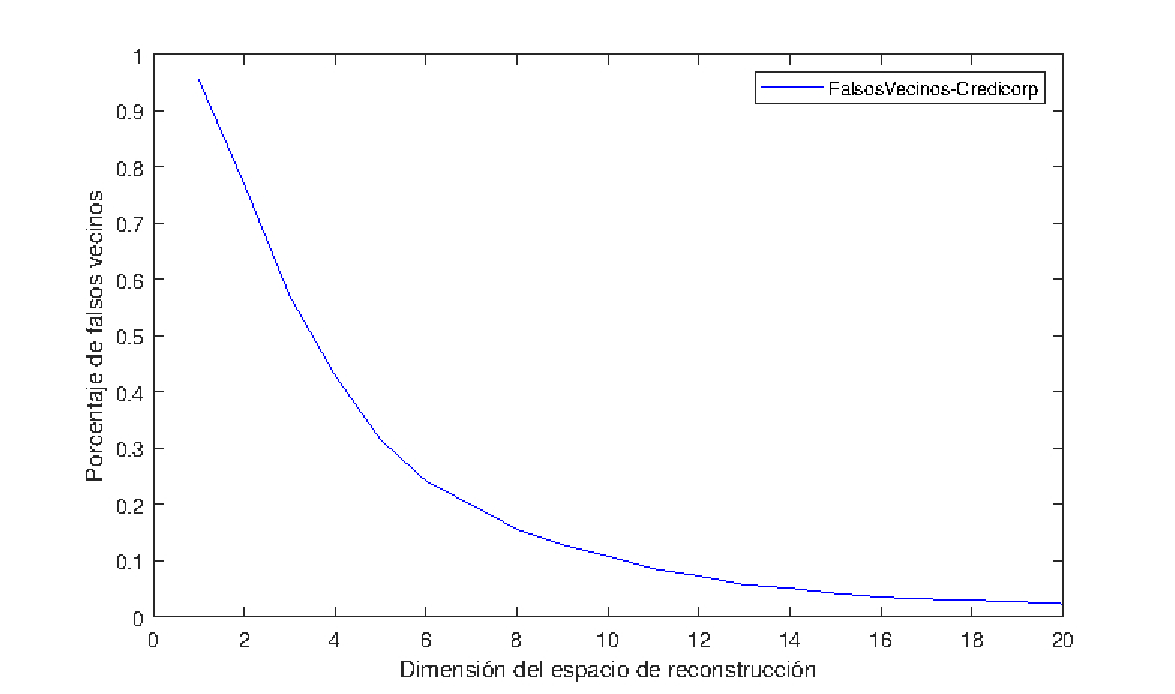}}
  \end{center}
 \caption{Embedding dimension $ n = $ 12 and $ n = $ 11, corresponding to the price series of the companies TelfBC and Credicorp.}
 \label{false}
 \end{figure}

For the reconstruction of the attractors, corresponding to the price series of the companies TelfBC and Credicorp, some parameters must be determined. The delay time is determined using the mutual information algorithm and the mergulho dimension is estimated using the false neighbors algorithm. \\
The time delay in rebuilding the dynamics of the price series for TelfBC was $ T = $ 9 and for Credicorp it was $ T = $ 6. These results are shown in the figure \ref{delay}.
\\
Embedding dimension is shown in figure \ref{false}. The percentage of false neighbors is almost zero when $ n = $ 12 for TelfBC and $ n = 11$ for Credicorp.
 
\begin{figure}[h]
  \begin{center}
  \scalebox{0.6}{\includegraphics{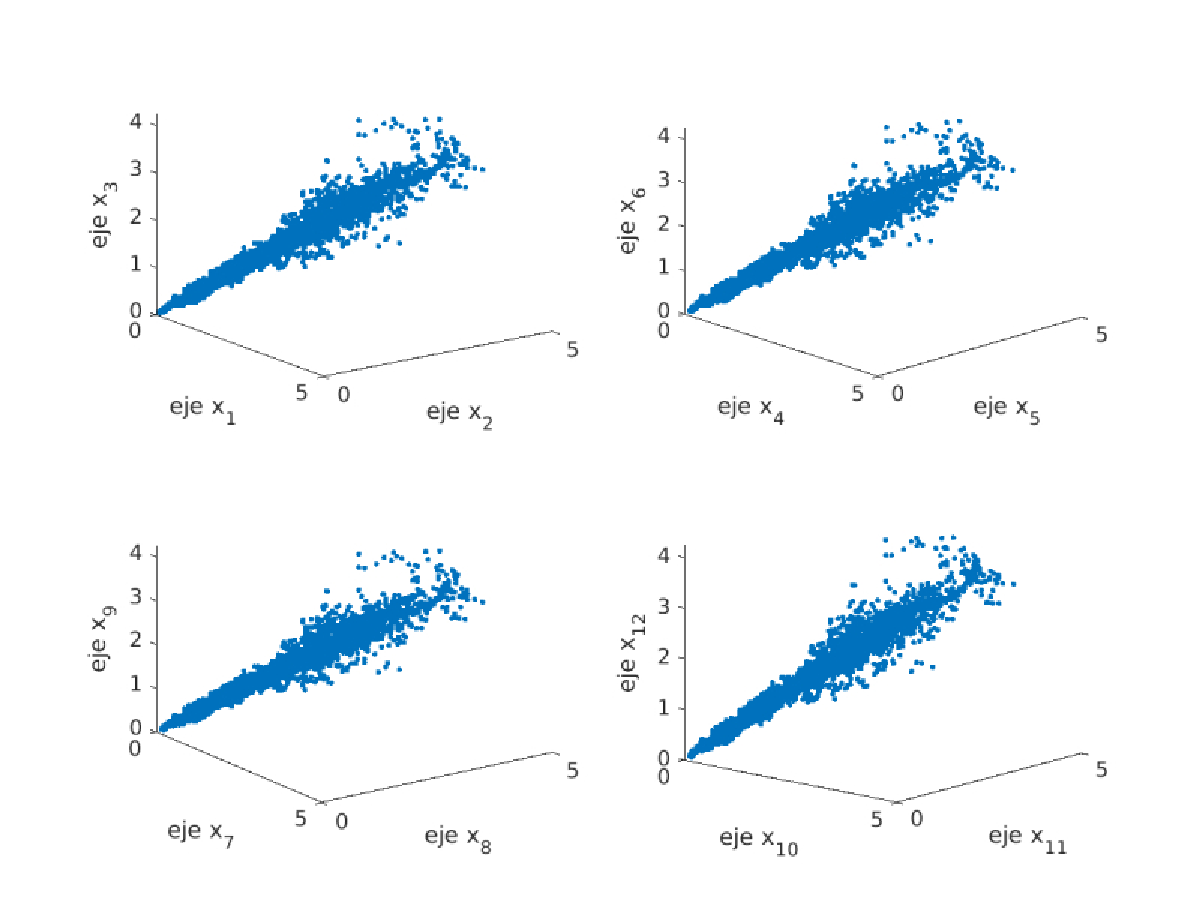}}
  \end{center}
 \caption{Reconstruction of attractors.}
 \label{atractorTelfBC}
 \end{figure}

Knowing the delay time $ T = 9$ and the mergulho dimension $ n = 12$, we apply the application of delay coordinates to reconstruct the hidden attractor in the series of prices of the shares of the company TelfBC. The figure \ref{atractorTelfBC} shows the projection of this attractor on the coordinate axes. 

\begin{figure}[h]
  \begin{center}
  \scalebox{0.6}{\includegraphics{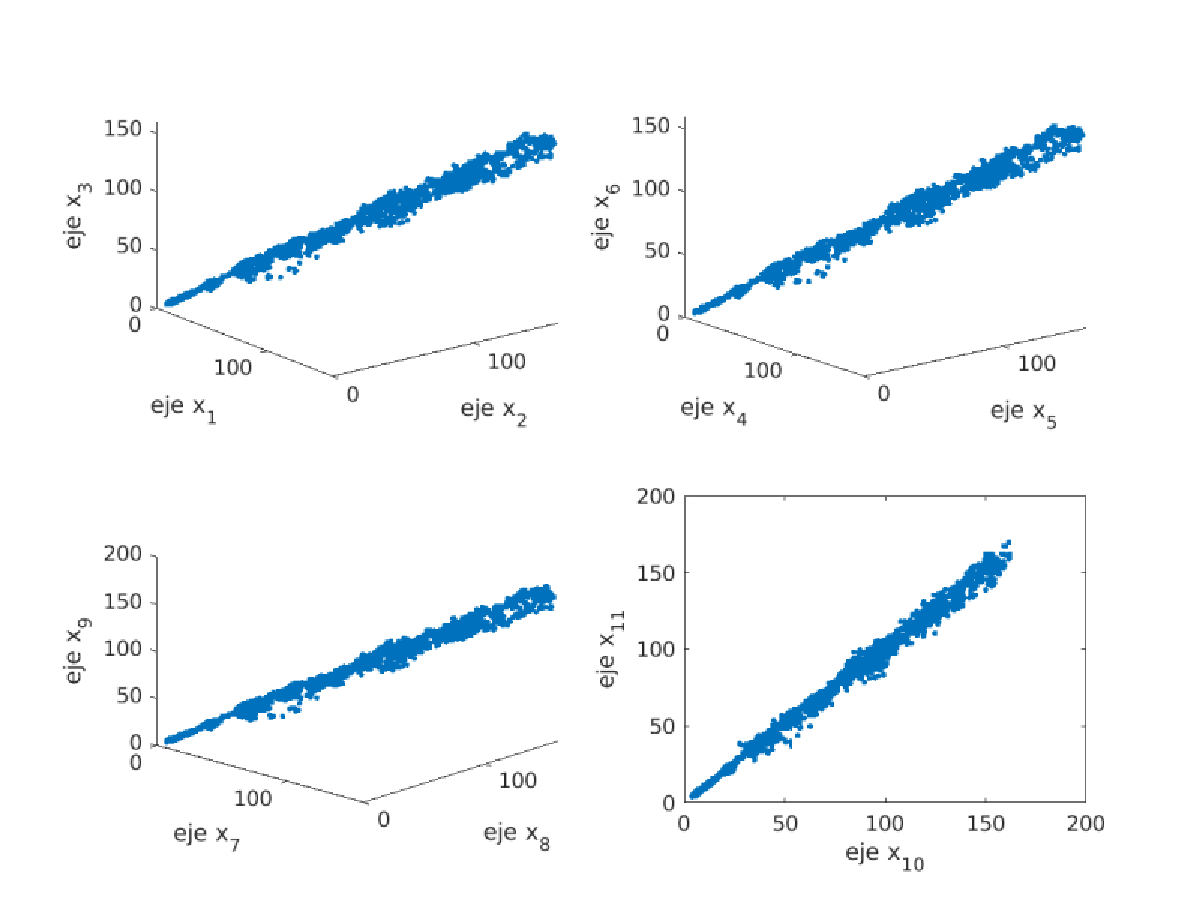}}
  \end{center}
 \caption{Reconstruction of attractors.}
 \label{atractorcred}
 \end{figure}

Knowing the delay time $ T = 6$ and the mergulho dimension $ n =11$, we apply the application of delay coordinates to reconstruct the hidden attractor in the series of prices of the shares of the company Credicorp. The figure \ref{atractorcred} shows the projection of this attractor on the coordinate axes.
\newpage
\subsection{Shannon entropy}
\begin{figure}[h]
  \begin{center}
  \scalebox{0.37}{\includegraphics{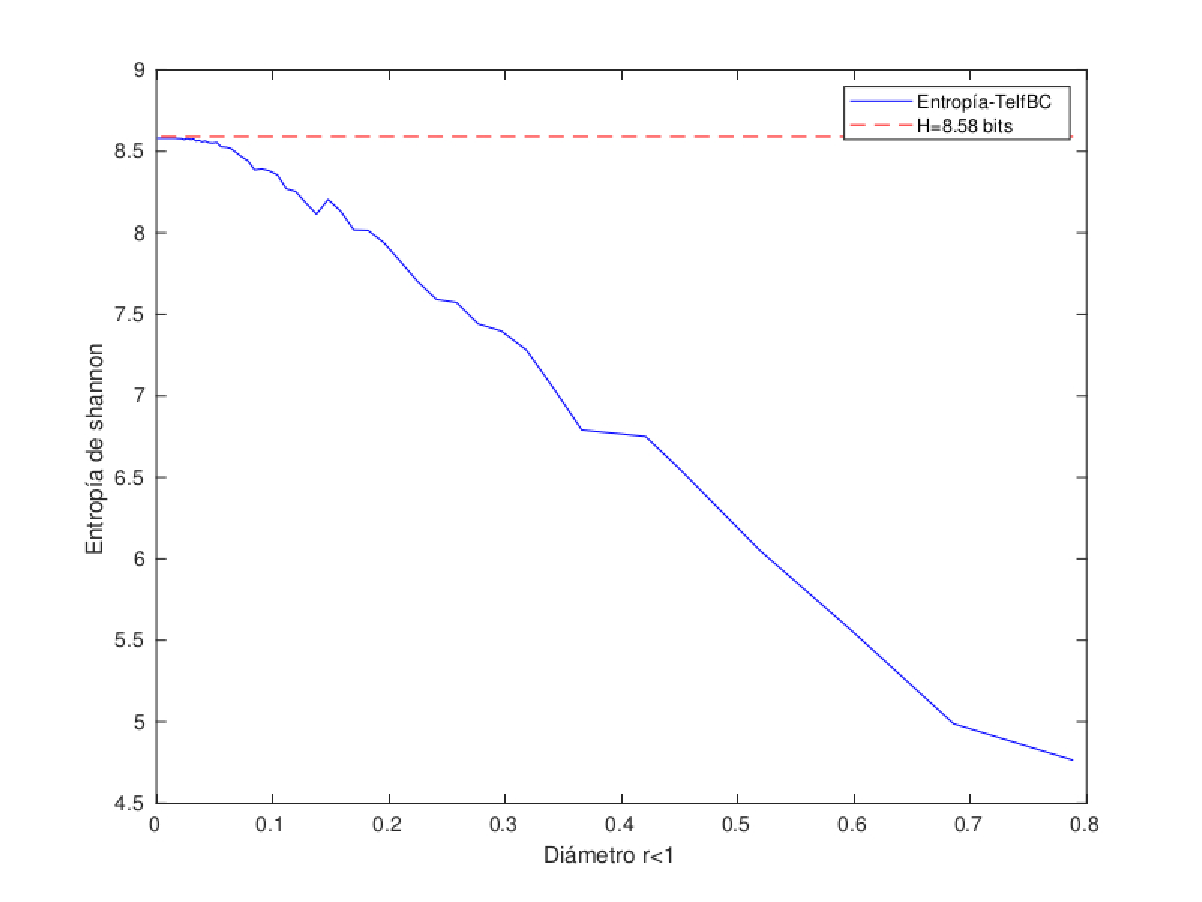}}
 \scalebox{0.37}{\includegraphics{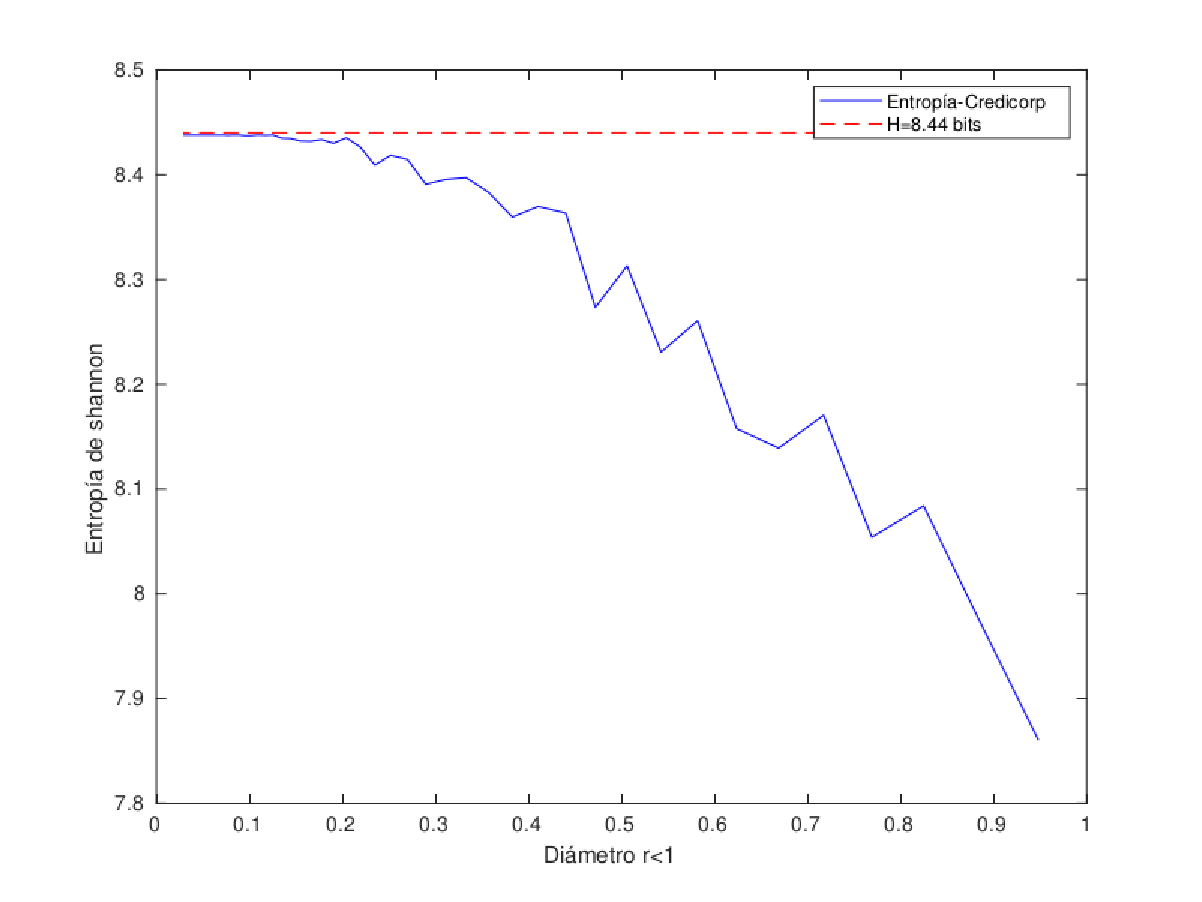}}
  \end{center}
 \caption{Shannon entropy $ H = 8.58$, 8.44 bits for the share price dynamics of the companies TelfBC and Credicorp.}
 \label{entropy}
 \end{figure}

Once the attractor $ A $ has been reconstructed via the delay coordinates, corresponding to the financial market price series, the Shannon entropy is calculated for the price dynamics of the shares of the companies TelfBC and Credicorp. For each of them the Shannon entropy is $8.58$ bits and $8.44$ bits of information. These values indicate the number of bits of information necessary to specify a point in the price dynamics of the shares of financial market companies. The figure \ref{entropy} shows these values.

\begin{figure}[h]
  \begin{center}
   \scalebox{0.37}{\includegraphics{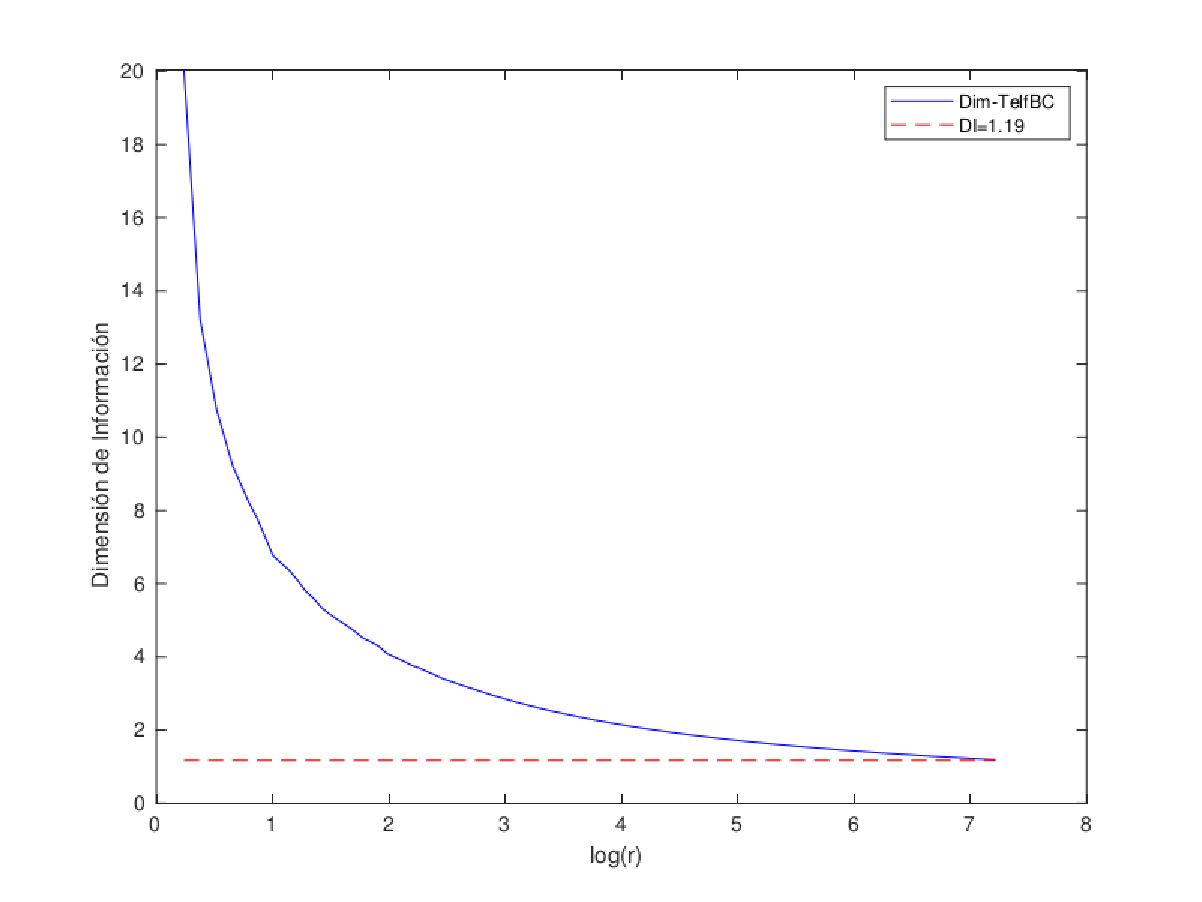}}
 \scalebox{0.37}{\includegraphics{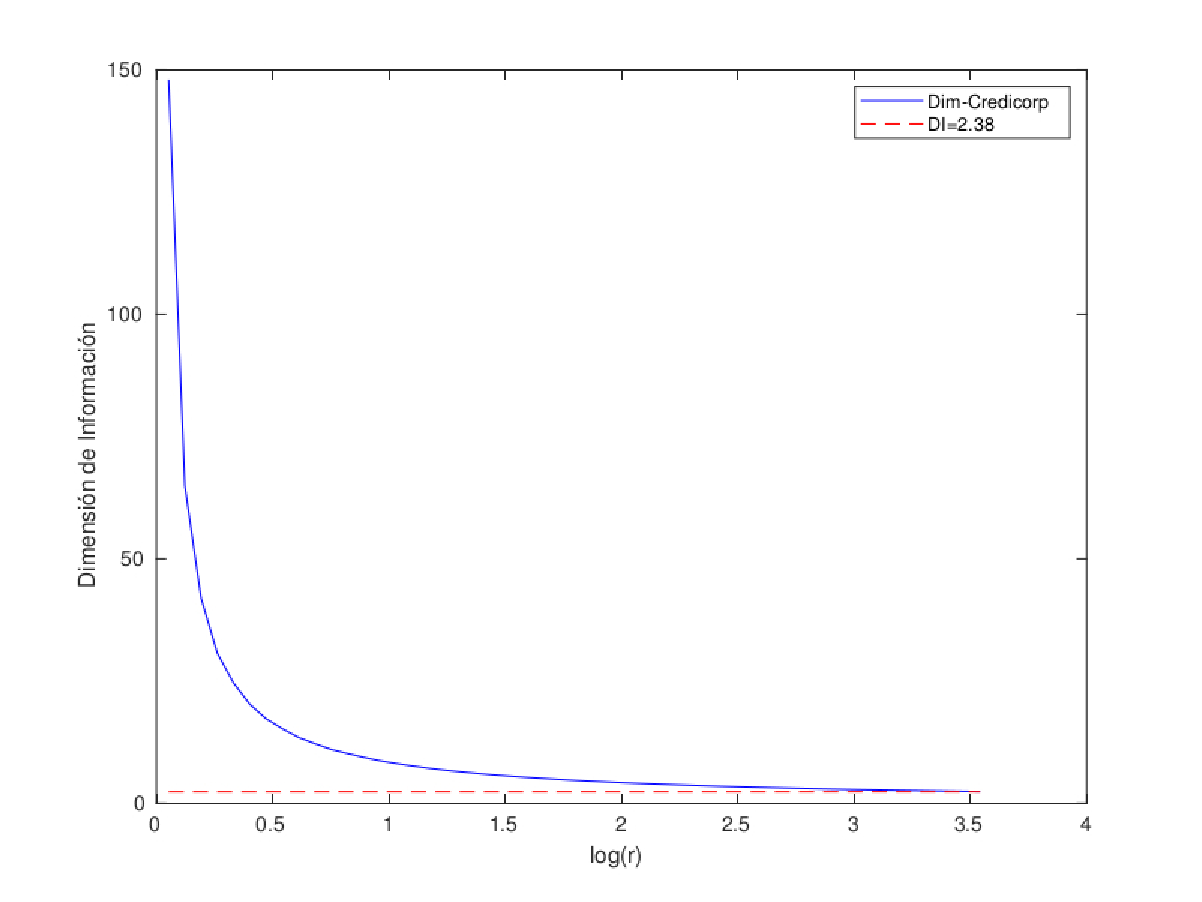}}
  \end{center}
 \caption{Information dimension $ DI = $ 1.19, 2.38 for the attractors corresponding to the price series of the companies TelfBC and Credicorp.}
 \label{diminfo}
 \end{figure}

Using the \ref {dim} theorem it is obtained that for the reconstructed attractor, for the price series of the company TelfBC, the information dimension is $ D_ {I} =1.19$ and for Credicorp it is $ D_ {I} = 2.38$. These values are shown in the figure \ref{diminfo}.
\\ 
These numerical values for the information dimension, 1.19 and 2.38, show the existence of a fractal set in the price dynamics of the shares of the companies TelfBC and Credicorp. In addition, these values quantify the complexity in the financial market, since it is necessary between 1 and 2 significant variables for TelfBC and between 2 and 3 significant variables for Credicorp to understand the dynamics of the prices of the shares of the mentioned companies.

\section{Conclusions}
Using the non-linear analysis and the information theory applied to the series of prices of the shares of the companies TelfBC and Credicorp, we obtained the following results:
\begin{enumerate}
\item The Shannon entropy for the share price dynamics of the companies TelfBC and Credicorp were 8.58 and 8.44 bits, respectively.
These indicated the number of bits of information necessary to specify a point in the price dynamics of the shares of financial market companies.
 \item The information dimension for the share price dynamics of financial market companies were 1.19 and 2.38, respectively for each company. The existence of a fractal attractor is evidenced in the dynamics
\end{enumerate}
\textcolor{green}{ORCID}\\
Jose Luis Ponte Bejarano \textcolor{cyan}{\url{https://orcid.org/0000-0002-4997-7950}},\\
Alexis Rodriguez Carranza \textcolor{cyan}{\url{https://orcid.org/0000-0002-0290-165X}},\\
Juan Carlos Ponte Bejarano \textcolor{cyan}{\url{https://orcid.org/0000-0002-8682-9682}}.\\
Segundo Eloy Soto Abanto \textcolor{cyan}{\url{https://orcid.org/0000-0003-1004-5520}}

\end{document}